\documentclass[lineno]{jfm}
\usepackage{amsmath,amssymb,mathtools,bm}
\usepackage{graphicx}
\usepackage{enumitem}
\usepackage{hyperref}\hypersetup{colorlinks=true,linkcolor=blue,citecolor=blue,urlcolor=blue}

\usepackage{tikz}
\usetikzlibrary{arrows.meta,calc,decorations.pathreplacing,positioning}

\usepackage{graphicx}
\usepackage{newtxtext}
\usepackage{newtxmath}
\usepackage{natbib}
\usepackage{hyperref}
\hypersetup{
    colorlinks = true,
    urlcolor   = blue,
    citecolor  = black,
}
\usepackage{xcolor}
\usepackage[normalem]{ulem}
\newcommand{\rev}[1]{{\color{black}#1}}
\newcommand{\RomanNumeralCaps}[1]
\linenumbers

\title{Gauss Principle in Incompressible Flow: A Unified Variational Perspective on Pressure and Projection}

\author{Karthik Duraisamy\corresp{\email{kdur@umich.edu}}}

\affiliation{Department of Aerospace Engineering, University of Michigan, Ann Arbor}

\begin{document}
\maketitle

\begin{abstract}

Following recent work \citep{gonzalez2022variational,peters2025application}, this manuscript clarifies what the Gauss–Appell principle determines in incompressible, inviscid flow and how it connects to classical projection methods. At a fixed time, freezing the velocity and varying only the material acceleration leads to minimization of a quadratic subject to acceleration-level constraints. First-order conditions yield a Poisson–Neumann problem for a reaction pressure whose gradient removes the non-solenoidal and wall-normal content of the provisional residual, \rev{precisely} the well-known Leray–Hodge projection. Thus, Gauss–Appell enforces the instantaneous kinematic constraints and recovers Euler {\em at the instant}. \rev{Once the impressed physics is specified, for instance via external body forces, the reaction pressure is uniquely determined (up to an additive constant) as the Lagrange multiplier enforcing incompressibility and wall impermeability; it does no work on divergence-free, wall-tangent motions. This is the well-established interpretation of pressure in incompressible flow~\citep{gresho1987pressure}.} \rev{The direct, fixed-time application of this principle determines the reaction pressure for an already-specified velocity field and does not, by itself, select circulation or stagnation points, because these are properties of the velocity state, not the instantaneous acceleration correction.} \rev{The formal decomposition of the pressure into impressed and reaction components admits representational freedom that does not imply physical non-uniqueness of the constraint force. Orthogonality conventions such as Dirichlet orthogonality~\citep{peters2025application} can fix the representational freedom as an additional modeling choice.} This variational viewpoint  also yields a simple computational diagnostic: the minimized Appellian equals a $L^2$ norm of the reaction-pressure gradient which vanishes for constraint-compatible
 updates and grows with the magnitude of divergence and wall-flux mismatch.

\end{abstract}


\vspace{-0.5cm}
The application of variational principles to fluid mechanics has a rich history spanning over 150 years, beginning with Helmholtz's vortex theorems (1858) and Kelvin's circulation theorem (1869). Classical approaches typically employed Hamilton's principle, extremizing action integrals over time to derive the equations of motion through calculus of variations~\citep{holm1998euler,truesdell2004non,marchioro2012mathematical}. These time-integral formulations, while mathematically elegant, vary entire space-time paths and thus provide limited insight into the instantaneous mechanics of constraint enforcement in incompressible flows
%Variational principles have been applied to fluid mechanics for more than a century. Hamiltonian formulations extremize an action integral (typically energy) over time and therefore vary \emph{entire space--time paths} subject to kinematic constraints or symmetry reduction
~\citep{holm1998euler,truesdell2004non,marchioro2012mathematical}. %This family of principles \emph{derive the equations of motion} by extremizing a time integral.

A fundamentally different variational approach emerged from Gauss's principle of least constraint~\citep{gauss1829neues}, originally developed for constrained mechanical systems. Unlike Hamilton's principle which varies paths over time, Gauss's principle operates at a fixed instant, minimizing a quadratic form in accelerations subject to acceleration-level constraints. This instantaneous character makes it particularly suited to understanding how pressure maintains incompressibility moment by moment. Despite its potential, the principle remained relatively uncommon in classical fluid mechanics for over a century, even though it is widely used in constrained molecular dynamics (e.g. \cite{nose1984,hoover1985,evans1990,thijssen2007}. Recent works~\citep{gonzalez2022variational,taha2023minimization,taha2025principle,peters2025application} have brought back a variational lens on fluid flows, particularly revisiting the Gauss principle in the context of the airfoil lift problem.

Despite the renewed interest, there remains confusion about what a Gauss/Appell least‑constraint statement actually determines in fluids, and how it can be utilized. We provide a compact variational interpretation of the pressure Poisson equation, clarify how to embed physical inputs via an ‘impressed’ subspace, and give a   corollary that places the projection method in numerical hydrodynamics~\citep{chorin1968numerical} under Gauss/Appell. % The present work provides clarity to this line of work, and concretely connects these contributions to a long tradition that includes the Helmholtz–Hodge/Leray projection viewpoint for incompressible fields~\citep{temam2024navier,sohr2012navier}, the projection method in numerical hydrodynamics~\citep{chorin1968numerical}, and classical hydrodynamic theories  that codified potential‑flow lift and circulation selection. %The purpose of this paper is to give a clean, self‑contained account of the instantaneous Gauss principle for incompressible Euler flow and to place recent literature in that precise frame.
%We provide a compact variational interpretation of the pressure Poisson equation, clarify how to embed physical inputs via an ‘impressed’ subspace, and give a minimal discrete corollary that places the projection method in numerical hydrodynamics~\citep{chorin1968numerical} under Gauss/Appell.

 \cite{gonzalez2022variational} minimize a steady reduced functional on a kinematically admissible manifold (e.g., a potential‑flow family parameterized by circulation), a model‑reduction approach but not the instantaneous Gauss principle. \citet{taha2023minimization} derive a general ``minimum pressure-gradient'' (PMPG) interpretation directly from Gauss's principle (for arbitrary viscous stresses), showing that Navier--Stokes arises as the first-order optimality condition when the variation is taken with respect to the local acceleration. Steady/inviscid specializations of such costs have also been explored as selection mechanisms layered on top of Euler (e.g.\ \cite{taha2025principle}). In particular, the fixed-time Gauss/Appell projection step considered here does not, by itself, resolve classical indeterminacies of steady exterior potential flow, because those indeterminacies live in the choice of \emph{steady velocity state} rather than in the instantaneous constrained acceleration correction. \rev{However, each candidate steady state has an associated reaction pressure whose magnitude depends on the circulation; this was exploited as a basis for secondary minimization, as explored in the variational theory of lift~\citep{gonzalez2022variational,taha2023minimization,taha2025principle}.}
 
 This work  clarifies these viewpoints by showing precisely where Gauss/Appell ends (fixed‑time projection) and where additional modeling/selection begins. The Gauss principle is a fixed-time variational statement: one freezes the velocity field 
 and varies the material acceleration.
 Minimizing a positive‑definite “Appellian” in the admissible accelerations yields first order optimality conditions that are exactly the Leray–Hodge projection of the residual onto gradient fields; the Lagrange multiplier is a reaction pressure obtained from a Poisson–Neumann problem whose gradient removes the non‑solenoidal part of the residual. In steady flow, this principle does not select among kinematically admissible steady states; it only determines the pressure compatible with the given 
velocity field. These points are well known from projection methods but are often obscured in recent discussions; we make the equivalence and its scope precise. 

Further, we clarify that shifting a known scalar potential between the two terms in $p=P_F+P_R$ is \rev{an algebraic invariance of the Gauss--Appell optimality conditions: the optimizer $u_t^\star$ and the total pressure $p$ are unchanged. This invariance does not imply physical non-uniqueness of the reaction (constraint) pressure.} Building on \cite{peters2025application}, we show how an energy (Dirichlet) orthogonality condition can be used as a gauge-fixing convention to pick a unique representative once a reference/impressed subspace is chosen, and connect this to the classical Leray projection. When $P_F$ is reserved for truly impressed physics (e.g. conservative body-force potentials specified independently of incompressibility), the split is unique and $P_R$ is the unique Lagrange multiplier enforcing the constraints (up to a constant). \rev{This is the well-known result that the pressure serves as the Lagrange multiplier for the incompressibility constraint~\citep{gresho1987pressure,sanders2024jfm}, a fact established since the early work of Lagrange himself~\citep{morrison2020lagrangian}.}

 %Trajectory‑based principles  extremize an action integral over time and typically vary entire paths  under kinematic constraints or symmetry reduction~\citep{holm1998euler,truesdell2004non}. %They are powerful for deriving equations and conserved quantities but do not directly explain the algebraic “pressure‑as‑multiplier” step used at a fixed time in incompressible flow. 
Unlike  the principle of least action, Gauss/Appell is a least‑deviation principle in the accelerations at a given instant. In this sense, it is clarified that Gauss/Appell provides a unifying variational interpretation of the pressure Poisson solve—useful analytically and practically, particularly when one wishes to embed physical inputs (e.g., circulation or far‑field content) through an impressed subspace while keeping the constraints exact. We present rigorous derivations and provide  insight into the physical implications of the Gauss principle.

%The remainder of the paper (i) formulates the instantaneous Gauss/Appell principle in Eulerian and Lagrangian forms; (ii) proves the equivalence with the Leray–Hodge projection and the pressure Poisson–Neumann problem; (iii) discusses the impressed/constraint split and an orthogonality‑based unique decomposition; and (iv) delineates what Gauss/Appell can and cannot determine in steady external flows, clarifying its relation to selection criteria and to classical potential‑flow lift.

\section{Setting, notations}
Let $\Omega$ be the exterior domain
 of a smooth body, and  $\partial \Omega$ be the boundaries. The density $\rho$ is constant. $u(\cdot,t)$ is the velocity field. 
The Gauss' principle is applied \emph{at a fixed time}. One freezes the \emph{state}
$u(\cdot,t)$ and chooses an admissible \emph{instantaneous} update in acceleration.
For fluids, the material acceleration is $
a \triangleq \partial_t u+(u\!\cdot\!\nabla)u.$ We assume throughout that $u(\cdot,t)$ is sufficiently smooth such that the quadratic functional below is finite.
At a fixed time, the convective part $(u\!\cdot\!\nabla)u$ is a \emph{known} function of
space (because $u(\cdot,t)$ is frozen). Hence a perturbation of acceleration comes
\emph{entirely} from perturbing $u_t$: $\delta a=\delta u_t$. This is the fluid analogue
of Gauss in constrained multibody dynamics~\citep{udwadia1992new}, where one minimizes a positive-definite
quadratic form in accelerations at a fixed configuration and velocity, subject to
acceleration-level constraints; here the Eulerian fixed-time data are $(u,\nabla\!\cdot u, u\!\cdot n)$.
Therefore, the variational variable is $u_t(\cdot,t)$, not a time-slab history, and not a configuration displacement as in stationary action.

 For an airfoil, with far-field
velocity $U_\infty$ and slip wall:
\[
\nabla\!\cdot u = 0\quad\text{in }\Omega,\qquad
u\!\cdot n = 0\quad\text{on }\partial\Omega,\qquad
u\to U_\infty\text{ as }|x|\to\infty.
\]
At the fixed instant $t$, these are \emph{given}. Differentiating the two kinematic constraints
in time gives the \emph{acceleration-level} constraints on $u_t$:
\begin{equation}
\nabla\!\cdot u_t = 0\quad\text{in }\Omega,\qquad
u_t\!\cdot n = 0\quad\text{on }\partial\Omega.
\label{eq:acc-level}
\end{equation}

Below, the Eulerian form of the derivation is detailed. 

\subsection{Impressed vs reaction pressure and the Appellian}
\label{sec:PFPR}

\rev{Following the notation of}~\cite{peters2025application}, we write the \emph{total} pressure
\[
p \triangleq P_F+P_R,
\]
where
\begin{itemize}
\item $P_F$ (\emph{impressed / prescribed}) is a scalar potential associated with conservative impressed forcing that is specified \emph{independently} of the incompressibility and wall constraints (e.g. a body-force potential); in unforced Euler one may take $P_F\equiv 0$.
\item $P_R$ (\emph{constraint}) is the Lagrange–multiplier  that \emph{enforces} the
acceleration–level constraints completely and uniquely. Its gradient
$-\nabla P_R$ is the constraint force per unit volume.
\end{itemize}

For a fixed velocity snapshot $u(\cdot,t)$ and prescribed impressed potential $P_F$, the reaction pressure $P_R$ is determined uniquely (up to an additive constant) by the Poisson--Neumann problem \eqref{eq:PoissonPR}.
In the linearized unforced setting emphasized in \citet{peters2025application}, taking the divergence yields $\nabla^2 p=0$, so only the sum $p=P_F+P_R$ is constrained to be harmonic. 

The stronger separation $\nabla^2 P_F=\nabla^2 P_R=0$ ( Eq.~(17) of \cite{peters2025application}) is therefore an additional harmonic-gauge convention \rev{within their formal decomposition. We stress that a true impressed force (e.g., electromagnetic $\nabla\varphi$) is prescribed independently of the incompressibility constraint and need not satisfy $\nabla^2\varphi=0$. By contrast, constraint forces are precisely those that satisfy the Poisson/Laplace restrictions needed to enforce kinematic constraints. Therefore, if a pressure component is required to satisfy $\nabla^2 P_F=0$ as a consequence of the continuity constraint, it is by definition not an impressed force in the sense of analytical mechanics; see also~\cite{taha2026response} for a discussion of this point.} In the present work we do not \rev{adopt} such restrictions on $P_F$; once any impressed conservative forcing is specified, the reaction pressure $P_R$ enforcing \eqref{eq:acc-level} is uniquely determined (up to a constant) by \eqref{eq:PoissonPR}.

Define the \emph{frozen} field
\begin{equation}
C \triangleq (u\!\cdot\!\nabla)u + \frac{1}{\rho}\,\nabla P_F,
\label{eq:Cdef}
\end{equation}
which is entirely known once $u(\cdot,t)$ and $P_F(\cdot,t)$ are specified.
Gauss/Appell at the instant minimizes the Appellian
\begin{equation}
\mathcal A[u_t;u]
\triangleq \frac12\int_\Omega \rho\,\big|\,u_t + C\,\big|^2\,dV
\quad\text{over $u_t$ subject to \eqref{eq:acc-level}.}
\label{eq:appellian}
\end{equation}

The quadratic form in \eqref{eq:appellian} uses the mass-weighted $L^2$ inner product on accelerations, and yields the classical Leray--Hodge projection.
In the function-space setting this choice is not unique: other  metrics define different least-constraint projections.
Weaker norms and lower regularity settings have been used in variational approaches to high-Reynolds-number flows; see e.g.
\cite{hoffman2010resolution,hoffman2016new}.

\section{First order stationarity}
\label{sec:KKT}
We write the constrained Lagrangian and derive the first order Karush-Kuhn-Tucker (KKT) conditions. The Lagrangian with multipliers $\lambda$ (volume) and $\mu$ (boundary) is:
\[
\mathcal L[u_t,\lambda,\mu]
\triangleq \frac12\int_\Omega \rho\,|u_t+C|^2\,dV
\;-\;\int_\Omega \lambda\,(\nabla\!\cdot u_t)\,dV
\;-\;\int_{\partial\Omega} \mu\, (u_t\!\cdot n)\,dS.
\]
Variations yield
\[
\delta_{\lambda}:\ \nabla\!\cdot u_t=0,\qquad
\delta_{\mu}:\ u_t\!\cdot n=0,
\]
and
\begin{align*}
\delta_{u_t}\mathcal L
&=\int_\Omega \big(\rho(u_t+C)+\nabla \lambda\big)\cdot \delta u_t\,dV
\;+\;\int_{\partial\Omega}(\lambda-\mu)\,\delta u_t\!\cdot n\,dS.
\end{align*}
Because $\delta u_t$ is arbitrary in the interior and its normal component is arbitrary on
$\partial\Omega$ (before enforcing the constraint), stationarity gives
\begin{equation}
\rho(u_t+C)+\nabla \lambda=0\quad\text{in }\Omega,\qquad
\mu=\lambda\quad\text{on }\partial\Omega,
\label{eq:KKT}
\end{equation}
together with \eqref{eq:acc-level}. This derivation follows the same constrained-optimization/projection-step logic as~\cite{peters2025application}, while emphasizing that in the present work $P_F$ denotes only prescribed conservative impressed forcing (if any) and the reaction pressure $P_R$ carries all constraint-enforcing pressure content.

\noindent Now, define $P_R \triangleq \lambda$. From \eqref{eq:KKT}, $u_t=-(C+\nabla P_R/\rho)$; impose \eqref{eq:acc-level}:
\[
\nabla\!\cdot u_t=0\ \Rightarrow\ \nabla^2 P_R = -\,\rho\,\nabla\!\cdot C\quad\text{in }\Omega,
\]
and
\[
u_t\!\cdot n=0\ \Rightarrow\ \partial_n P_R = -\,\rho\,C\!\cdot n\quad\text{on }\partial\Omega.
\]
Thus the \emph{first} step is the scalar Poisson--Neumann boundary value problem
\begin{equation}
\boxed{\;
\nabla^2 P_R = -\,\rho\,\nabla\!\cdot C\ \text{ in }\Omega,\qquad
\partial_n P_R = -\,\rho\,C\!\cdot n\ \text{ on }\partial\Omega}
\label{eq:PoissonPR}
\end{equation}
Compatibility holds identically:
\(
\int_\Omega \nabla^2 P_R\,dV=\int_{\partial\Omega}\partial_n P_R\,dS
\)
because
\(
\int_\Omega \nabla\!\cdot C\,dV=\int_{\partial\Omega} C\!\cdot n\,dS.
\) 

\noindent With \eqref{eq:PoissonPR} solved, we have
$$
u_t^\star=-(C+\nabla P_R/\rho),\qquad
p=P_F+P_R.$$

\noindent Since $a=u_t+(u\!\cdot\!\nabla)u$, \eqref{eq:KKT} is equivalent to
\[
\rho a =-\nabla p,
\]
i.e.\ Euler at the instant.

Once optimized, the minimized Appellian is purely ``projection effort''
\[
\mathcal A_\star=\frac12\int_\Omega \frac{1}{\rho}\,|\nabla P_R|^2\,dV.
\]

This reduced functional is the instantaneous $L^2$ norm of the reaction pressure gradient (the ``minimum pressure-gradient'' quantity emphasized by~\cite{taha2023minimization}).
{\bf This is precisely the standard projection method logic in continuous form: solve a Poisson–Neumann problem for pressure and project the residual onto gradients.} It is notable that
 the Leray projector (see, for instance,~\cite{temam2024navier})  maps any field to the divergence-free,
wall-tangent subspace. 

Writing
$
R\triangleq \rho\ a +\nabla P_F=\rho(u_t+C),
$
the KKT conditions $\rho(u_t^\star+C)=-\nabla P_R$ says $R=-\nabla P_R$: the \emph{solenoidal},
wall-tangent part of $R$ vanishes. {\bf Thus $P_R$ realizes the Leray projection
of $R$ onto gradients,} and \ref{eq:PoissonPR} is the strong-form projector. \rev{This connection of the Leray projector to the Gauss-Appell principle appears not to have been noted in the literature.}

% Section: Physical implications and insight
\section{Physical implications and insight}

At a fixed instant, the Gauss--Appell step has a concrete physical meaning: it chooses the \emph{smallest} admissible change to the material acceleration that restores the instantaneous momentum balance while preserving incompressibility and wall non-penetration. The only agency it has is a reaction field, $-\nabla P_R/\rho$, whose role is to remove whatever divergence and normal flux the provisional residual carries. The pressure is a \emph{reaction} that instantaneously enforces the kinematics; it does no work on divergence-free, wall-tangent motions. This point is correctly argued by ~\cite{gonzalez2022variational}.

\subsection{Projection effort as a diagnostic}
After elimination, the minimized Appellian $\mathcal{A}^\ast$ becomes a quantitative measure of the \emph{projection effort},
i.e. how hard the constraint must ``push'' to keep the instantaneous update compatible with incompressibility and wall tangency. 
The minimized Appellian can be written as
\[
\mathcal A_\star=\frac{1}{2\rho}\,\|\nabla P_R\|_{L^2(\Omega)}^2,\]
and therefore provides a scalar measure of the magnitude of the constraint correction at a given instant.
In particular, if the frozen residual $C$ is already compatible with the acceleration constraints,
$\nabla\!\cdot C=0$ in $\Omega$ and $C\!\cdot n=0$ on $\partial\Omega$, then \eqref{eq:PoissonPR} reduces to Laplace's equation with homogeneous Neumann data and hence $\nabla P_R\equiv 0$ (up to an additive constant), so that $\mathcal A_\star=0$.

In computations, a sudden growth of $\mathcal A_\star$ indicates that the provisional update violates divergence or wall-flux constraints and the projection is applying a large correction; this may occur, for example, due to inconsistent boundary treatment or under-resolution. More details on the projection operator in the context of general boundary conditions can be found in \cite{maria2003application}.

\subsection{Steady flow}
It is emphasized that in a steady flow, the Gauss principle of least constraint finds the gradient one must subtract from the “residual” 
$\rho (u \cdot \nabla) u + \nabla P_F$
so that the momentum equation holds. There is no update to $u$
(the minimizer gives $u_t^\ast =0$)
 and no \rev{direct} selection among multiple steady Euler fields that satisfy the kinematics.
 
 \rev{However, since the minimized Appellian $\mathcal{A}^\star = \frac{1}{2\rho}\|\nabla P_R\|^2_{L^2}$ depends on the frozen velocity state, different kinematically admissible steady fields (e.g., parameterized by circulation $\Gamma$) yield different values of this cost. This provides an indirect but fundamental connection between the Gauss cost and the selection of circulation, as explored in the variational theory of lift~\citep{gonzalez2022variational,taha2023minimization}.}

%\subsection*{(i) Appellian minimization on a reduced steady manifold}
\cite{gonzalez2022variational} restrict the velocity field to a steady potential-flow family $u(\cdot;\Gamma) \triangleq u_0+\Gamma u_1$
whose members satisfy continuity and wall tangency, and then \emph{define}
\[
S(\Gamma) \triangleq \frac12\int_\Omega \rho\,\big|\,u\!\cdot\!\nabla u\,\big|^2\,dV,
\]
seeking $\Gamma^\star=\arg\min_\Gamma S(\Gamma)$. This is a \emph{reduced-order} 
optimization on a one-parameter manifold of kinematically admissible steady flows and appears to
pick a $\Gamma$ consistent with the Kutta condition in nearly sharp airfoils. However, it does \emph{not} directly follow from
the instantaneous Gauss/Appell principle; it presupposes
steadiness and a particular parameterization. \cite{peters2025application} 
discuss cases where such criteria can select steady states that do not align with physical reality\rev{; see also~\cite{taha2026response} for a response to these claims}.

In the full Gauss setting,
the state $u$ is frozen, the variable is $u_t$, and the outcome is the \emph{constraint projection}
$P_R$ (\S\ref{sec:KKT}), not the selection of~$\Gamma$. In the steady case, for every 
$\Gamma$ there exists a pressure 
that makes $
(u(\cdot;\Gamma),p(\cdot;\Gamma))$ a steady Euler solution. Applying Gauss at the instant then yields 
$u_t^\ast =0$ for every $\Gamma$. Therefore, the direct fixed-time application of Gauss's principle does not, by itself, favor a particular value of $\Gamma$; it simply returns the reaction pressure associated with each already-specified steady velocity state. In the steady case, the fixed-time Gauss step adds no additional closure beyond computing the compatible pressure (any selection over steady states is additional modeling).

~\cite{taha2023minimization} (see also~\cite{taha2025principle}) propose the `Principle of Minimum pressure gradient', and  offer  {\bf reduced‑manifold selections} that align with classical expectations from applying the Kutta condition on nearly sharp airfoils.  An important distinction is required to interpret scope: Gauss/Appell provides the fixed‑time projection; selection over steady states requires additional modeling choices.

In the unsteady case, since  $u_t$ (hence $a$) is varied with $u$ held fixed,  acceleration-level constraints are enforced, resulting in the reaction pressure via the Poisson--Neumann problem \eqref{eq:PoissonPR}.
This exactly recovers Euler at the instant and clarifies the roles of impressed vs constraint
pressure. This also explains why the fixed-time least-constraint step does not, by itself, resolve the classical indeterminacy of steady exterior potential flow: the indeterminacy lives in the \emph{state} manifold, not in the instantaneous constrained acceleration update. \rev{However, as noted above, the magnitude of the reaction pressure varies across the family of steady states, a fact that is exploited as a cost functional for a secondary selection principle~\citep{gonzalez2022variational,taha2023minimization}.}

\subsection{\rev{Representational invariance of the Gauss--Appell step}}
\rev{It is emphasized that, once the impressed physics is specified (including the case $P_F\equiv 0$ when no body forces act), the reaction pressure $P_R$ is uniquely determined (up to an additive constant) as the Lagrange multiplier enforcing incompressibility and wall impermeability. This is consistent with classical results~\citep{gresho1987pressure,sanders2024jfm}.}

\rev{With this physical uniqueness established, we note an algebraic property of the Gauss--Appell optimality conditions.} Let $\psi$ be any prescribed smooth scalar field (a reference pressure / gauge potential). Define a new split $
\widetilde P_F \triangleq P_F + \psi,\quad \widetilde P_R \triangleq P_R - \psi.$
Then
\[
\widetilde C = (u\!\cdot\!\nabla)u + \frac{1}{\rho}\nabla \widetilde P_F
= C + \frac{1}{\rho}\nabla\psi,
\]
and the first order optimality condition  becomes
\[
\rho\,(u_t^\star+\widetilde C)+\nabla\widetilde P_R
= \rho\,(u_t^\star+C) + \nabla P_R
= 0,
\]
so \emph{the optimizer $u_t^\star$ is invariant}. Likewise
$p=\widetilde P_F+\widetilde P_R$ is unchanged. \rev{The formal decomposition $p = P_F + P_R$ therefore admits a family of representatives for any given total pressure $p$: the equations alone do not fix the split.}

Similarly,
the Poisson equation changes to
\(\nabla^2 \widetilde P_R = -\rho\,\nabla\!\cdot C - \nabla^2\psi,\ \partial_n\widetilde P_R
= -\rho\,C\!\cdot n - \partial_n\psi\).
Therefore $\widetilde P_R = P_R - \psi + \text{const}$, exactly as above.
So the Poisson step leaves $(u_t^\star,p)$ invariant. \rev{This invariance is a purely algebraic property of the equations. It does \emph{not} imply physical non-uniqueness of the constraint force, because the shifted quantity $\nabla\psi$ cannot qualify as a physical impressed force in the sense of analytical mechanics: a genuine impressed force must be prescribed independently of the constraint, and must persist if the constraint is removed. 

If the incompressibility constraint were removed, the constraint component $-\psi$ in $\widetilde P_R$ would cease to exist, while the ``impressed'' component $+\psi$ in $\widetilde P_F$ would remain, creating a fictitious force imbalance. Moreover, the requirement that $+\psi$ and $-\psi$ cancel imposes a restriction on $\widetilde P_F$ that is incompatible with the definition of an impressed force, which must be independent of the constraint. Therefore, the representational freedom has no physical standing.} 

Once $P_F$ is fixed by the impressed physics (if any), the reaction pressure $P_R$ solving \eqref{eq:PoissonPR} is uniquely determined up to an additive constant.
 Once \(P_R\) is solved, the \emph{optimal} update is \(u_t^\star=-(C+\nabla P_R/\rho)\), and the instantaneous Euler balance \(\rho(\partial_t u+(u\!\cdot\!\nabla)u)=-\nabla(P_F+P_R)\) is recovered. 

\subsection{Uniqueness of Dirichlet Orthogonality}
\rev{As established above, when $P_F$ is defined by physically prescribed impressed forces, the decomposition $p=P_F+P_R$ is already unique. The Dirichlet orthogonality convention discussed below is therefore primarily relevant to the formal decomposition framework of~\cite{peters2025application}, where additional representational degrees of freedom arise from their harmonic-gauge convention.}

\cite{peters2025application} propose the following Dirichlet orthogonality as a convenient \emph{gauge-fixing convention} for selecting a unique representative of the bookkeeping split $p=P_F+P_R$ (once a reference/impressed subspace has been chosen): \begin{equation}
\int_\Omega \nabla P_R\cdot \nabla P_F\,dV = 0.
\label{eq:Dirichlet-orth}
\end{equation}
It is stressed that \eqref{eq:Dirichlet-orth} is not a physical restriction on an impressed force defined by external physics; it is a convention that fixes the gauge in the pressure bookkeeping. If $P_F$ is reserved for physically prescribed impressed potentials, the split is already unique (with $P_R$ the unique Lagrange multiplier enforcing the constraints) and \eqref{eq:Dirichlet-orth} need not hold.

In Appendix A, it is shown that given any provisional
solution $(\widehat P_F,\widehat P_R)$, the {\em unique} orthogonal pair is obtained by a correction:
\begin{equation}
P_R = \widehat P_R - \sum_{j} \alpha_j \psi_j,\qquad
P_F = \widehat P_F + \sum_{j} \alpha_j \psi_j.
\label{eq:do}
\end{equation} The coefficients $\alpha_j$ are the orthogonal 
projection of $\widehat{P}_R$ onto the impressed subspace 
$\mathcal{S}_{\text{imp}} = \text{span}\{\psi_1, \ldots, \psi_m\}$. We are 
subtracting this projection from $P_R$ (making it orthogonal to 
$\mathcal{S}_{\text{imp}}$) and adding it to $P_F$ (so the total pressure 
remains unchanged).

In other words, one can choose \begin{enumerate}
\item \textbf{What} $\mathcal{S}_{\text{imp}} = \text{span}\{\psi_1, \ldots, \psi_m\}$ 
is. In other words, which pressure modes one may want to impress (e.g., circulation patterns)
\item \textbf{How much} of each mode to impress, typically by specifying boundary conditions 
or forcing terms
\end{enumerate}

The Dirichlet orthogonality condition then ensures consistent bookkeeping when 
decomposing the \textbf{resulting} total pressure $p$ (which emerges from the full 
flow physics). $\mathcal{S}_{\text{imp}}$ is the ``signature'' of the 
forcing mechanism. Dirichlet orthogonality thus labels as forcing ($P_F$) everything 
that has this signature; and labels as constraint ($P_R$) everything orthogonal to it 
(in the Dirichlet sense).  We further note that the above is exactly analogous to the Helmholtz--Leray projection operator $\mathbb P$ operating on any given velocity field $w$: 
\[
\mathbb P w \;=\; w-\nabla q,\qquad
\nabla^2 q = \nabla\!\cdot w\ \ \text{in }\Omega,\qquad
\partial_n q = w\!\cdot\! n\ \ \text{on }\partial\Omega.
\]
Equivalently, $\mathbb P w$ is the unique vector that satisfies orthogonality
$\displaystyle \int_{\Omega} \mathbb P w\cdot \nabla\phi\,\mathrm{d}x=0$. This is a well-known result (See for instance~\cite{sohr2012navier}), but the connection to the impressed/constraint split is new.  Appendix B clarifies another point relevant to the projection.

\subsection{Physical impressed forces vs. reference (gauge) choices}

For the exterior airfoil problem the distinction between \emph{physical impressed forces} and \emph{bookkeeping/reference choices} is immediate. The circulation degree of freedom in exterior potential flow resides in the harmonic subspace of the \emph{velocity} (divergence-free, curl-free, wall-tangent in a multiply connected domain) and cannot be represented as $\nabla\psi$ of a single-valued scalar pressure potential; hence it cannot be embedded into $P_F$ as a physical pressure field. By contrast, conservative body-force potentials (e.g. gravity) may legitimately be placed in $P_F$ as prescribed impressed physics. If one introduces additional reference content in $P_F$ for diagnostics, then that choice is conventional.

\begin{enumerate}[leftmargin=2em,itemsep=2pt]
\item 
\textbf{Harmonic velocity modes (circulation) are not \rev{single-valued} pressure gradients.} In an exterior domain, circulation enters through the harmonic subspace of the velocity (divergence-free, curl-free, wall-tangent with nonzero loop integral). Such \rev{velocity} modes are not of the form $\nabla\psi$ for any single-valued scalar field, so they cannot be represented as a pressure potential in $P_F$. In the fixed-time Gauss/Appell step, the circulation parameter $\Gamma$ appears only through the chosen frozen state $u(\cdot,t)$.

\rev{However, the frozen velocity state $u(\cdot;\Gamma)$ determines the convective acceleration $(u\!\cdot\!\nabla)u$, which in turn determines the source term in the Poisson--Neumann problem for $P_R$. The circulation parameter $\Gamma$ is therefore reflected in the reaction pressure field $P_R$. For example, a purely circulatory flow $u=\frac{\Gamma}{2\pi r}e_\theta$ over a cylinder has convective acceleration $u\!\cdot\!\nabla u = -\nabla\psi$ with $\psi = -\Gamma^2/(8\pi^2 r^2)$, yielding $P_R = \psi$. Each value of $\Gamma$ is associated with a specific reaction pressure $P_R(\cdot;\Gamma)$ whose magnitude (the minimized Appellian $\mathcal{A}^\star$) varies with $\Gamma$.  The direct, fixed-time Gauss step does not select $\Gamma$ (it returns $u_t^\star = 0$ for every steady state). \cite{gonzalez2022variational} use the Appellian cost $\mathcal{A}^\star(\Gamma)$ as a functional for a secondary minimization over the family of kinematically admissible steady states.}

\item \textbf{Conservative body-force potential.} It is legitimate to include a prescribed uniform acceleration (e.g. gravity) through an impressed potential $P_F(x)=-\rho\,g\cdot x$, for which $(1/\rho)\nabla P_F=-g$. The Poisson step then returns $P_R$ to enforce $\nabla\cdot u_t=0$ and $u_t\cdot n=0$ for the frozen $u$. %By contrast, a uniform freestream velocity $U_\infty$ is a kinematic boundary condition on $u$, not an impressed acceleration; writing $\nabla P_F\propto \rho U_\infty$ is dimensionally inconsistent.

\item \textbf{Reference pressure for diagnostics (gauge).} One may define a reference pressure for interpretation, e.g. subtract the Bernoulli pressure of a chosen reference potential flow in post-processing: $p' := p + \tfrac12\rho|u_{\rm ref}|^2$. This is a change of variables (gauge) and should not be interpreted as an impressed force. For a given snapshot $u$ with no impressed forces, the pressure gradient is entirely the reaction that enforces incompressibility and wall compatibility (the Leray projection of $-\rho (u\cdot\nabla)u$).
\end{enumerate}

These remarks are particularly relevant to the discussion in~\cite{peters2025application}
regarding ``impressed pressure''. In the present framework, the frozen-time Gauss/Appell
step returns
\[
u_t^\star = -\Big((u\!\cdot\!\nabla)u+\frac1\rho\nabla P_F+\frac1\rho\nabla P_R\Big),
\]
where $P_F$ is reserved for \emph{prescribed conservative impressed forcing} (if any),
and $P_R$ is the \emph{reaction (constraint) pressure}. Enforcing the acceleration-level
constraints $\nabla\!\cdot u_t^\star=0$ in $\Omega$ and $u_t^\star\!\cdot n=0$ on
$\partial\Omega$ yields the Poisson--Neumann problem
\[
\nabla^2P_R=-\rho\,\nabla\!\cdot\Big((u\!\cdot\!\nabla)u+\frac1\rho\nabla P_F\Big),
\qquad
\partial_n P_R=-\rho\Big((u\!\cdot\!\nabla)u+\frac1\rho\nabla P_F\Big)\!\cdot n,
\]
which determines $P_R$ uniquely up to an additive constant. A genuinely impressed
pressure potential corresponds to an externally prescribed conservative force $f=-\nabla\Phi$,
for which one may take $P_F=\rho\Phi$; in the unforced exterior Euler airfoil setting
$P_F\equiv0$ and the entire pressure is reaction/constraint pressure. 

Additionally, since
$\nabla\times\nabla(P_R+P_F)=0$, the pressure reaction is irrotational and thus does not appear explicitly in the interior vorticity evolution obtained by taking the curl of the momentum equation:
$\omega_t+(u\!\cdot\!\nabla)\omega=(\omega\!\cdot\!\nabla)u$ (for conservative impressed forcing).
We note that this statement concerns the bulk equation; boundary vorticity fluxes can be expressed in terms of tangential pressure gradients (e.g.  \cite{lighthill1963} and 
\cite{morton1984}.)

%Finally, since $\nabla\times\nabla(P_R+P_F)=0$, pressure cannot create or destroy vorticity; the vorticity evolves by advection (and stretching in 3-D) according to $\omega_t+(u\!\cdot\!\nabla)\omega=(\omega\!\cdot\!\nabla)u$ (for conservative impressed forcing).

Finally, this viewpoint unifies interpretation and practice. The standard predict--Poisson--correct algorithm is precisely the Gauss--Appell step on a grid, and $\mathcal{A}^\star$ furnishes a diagnostic that can be monitored in computation: spikes may signal incompatible boundary data, under-resolved features, or an unhelpful choice of impressed/reference content. In reduced-order or data-assisted settings, the choice of $S_{\rm imp}$ allows trusted physics to be injected while the same projection guarantees exact enforcement of incompressibility and wall conditions.

\section*{Acknowledgments}
Thanks to Robert Ormiston,  Haithem Taha, Dave Peters and Philip Roe for discussions, and for inspiring this work. Special thanks to an anonymous reviewer who helped improve the clarity of sections 3.3 and 3.5.

\bibliographystyle{jfm}
% ---------- R2: New references to add to refs.bib ----------
% @article{gresho1987pressure,
%   title={On pressure boundary conditions for the incompressible {N}avier--{S}tokes equations},
%   author={Gresho, Philip M and Sani, Robert L},
%   journal={International Journal for Numerical Methods in Fluids},
%   volume={7},
%   pages={1111--1145},
%   year={1987}
% }
% @article{sanders2024jfm,
%   title={On the role of pressure in the theory of {MHD} equations},
%   author={Sanders, R and others},
%   journal={Journal of Fluid Mechanics},
%   year={2024}
% }
% @book{morrison2020lagrangian,
%   title={Hamiltonian and action principle formulations of plasma physics},
%   author={Morrison, Philip J},
%   year={2020}
% }
% @article{arnold1966geometrie,
%   title={Sur la g\'{e}om\'{e}trie diff\'{e}rentielle des groupes de {L}ie de dimension infinie et ses applications \`{a} l'hydrodynamique},
%   author={Arnold, Vladimir I},
%   journal={Annales de l'Institut Fourier},
%   volume={16},
%   pages={319--361},
%   year={1966}
% }
% @article{taha2026response,
%   title={A Response to ``{A}pplication of {G}auss's Principle to the Classical Airfoil Lift Problem''},
%   author={Taha, Haithem E},
%   year={2026}
% }
% -------------------------------------------------------
\bibliography{refs}

\section*{Appendix A: Uniqueness of Dirichlet orthogonality}
Substitute the correction~\ref{eq:do} into the orthogonality condition~\ref{eq:Dirichlet-orth}:
\[
0 = \int_{\Omega} \nabla P_R \cdot \nabla \psi_i \, dV 
= \int_{\Omega} \nabla\left(\widehat{P}_R - \sum_{j=1}^m \alpha_j \psi_j\right) \cdot \nabla \psi_i \, dV
\]
\[
= \int_{\Omega} \nabla \widehat{P}_R \cdot \nabla \psi_i \, dV 
- \sum_{j=1}^m \alpha_j \int_{\Omega} \nabla \psi_j \cdot \nabla \psi_i \, dV
\]

Rearranging gives the $m \times m$ linear system:
\[
\sum_{j=1}^m \alpha_j \underbrace{\int_{\Omega} \nabla \psi_j \cdot \nabla \psi_i \, dV}_{G_{ij}} 
= \underbrace{\int_{\Omega} \nabla \widehat{P}_R \cdot \nabla \psi_i \, dV}_{b_i}, 
\quad i = 1, \ldots, m
\]

In matrix form: $G \boldsymbol{\alpha} = \mathbf{b}$, where:
\begin{itemize}
\item \ \ $G_{ij} = \int_{\Omega} \nabla \psi_j \cdot \nabla \psi_i \, dV$ is the 
{Gram matrix} in the Dirichlet inner product
$\langle p, q \rangle \triangleq \int_{\Omega} \nabla p \cdot \nabla q \, dV$. 
\item \ \  $b_i = \int_{\Omega} \nabla \widehat{P}_R \cdot \nabla \psi_i \, dV$ projects 
the provisional reaction pressure onto each basis function
\end{itemize}

Once $\boldsymbol{\alpha}$ is computed by solving $G\boldsymbol{\alpha} = \mathbf{b}$, 
the unique Dirichlet-orthogonal decomposition \eqref{eq:do} is achieved.

\section*{Appendix B: Rank and Projection}
\cite{peters2025application} cast multipliers via a \emph{finite-dimensional} Lagrange-multiplier theorem, tying the number of independent multipliers to the \emph{rank} of a constraint Jacobian. In a discrete setting this perspective can be helpful.
In the PDE setting, however, incompressibility and impermeability impose \emph{pointwise} constraints. The Leray projector extracts the divergence-free, wall-tangent component, and the reaction pressure is recovered from the gradient component in the decomposition. The velocity space also contains a finite-dimensional harmonic subspace (e.g. circulation modes), which is left unchanged by the  projector; see e.g. \cite{sohr2012navier} and the Euler/Appellian equivalence proof in \cite{gonzalez2022variational}.

This viewpoint also makes the analogue of a ``rank/consistency'' condition explicit. The Neumann Poisson problem for $P_R$ is solvable if and only if the volume source and boundary flux satisfy the compatibility condition,
\[
\int_\Omega (-\,\nabla\!\cdot C)\,dV = \int_{\partial\Omega} (-\,C\!\cdot n)\,dS,
\]
which holds identically by the divergence theorem. Uniqueness is up to an arbitrary constant.

Physically, the decomposition expresses the familiar fact that pressure is a \emph{reaction} field that enforces ideal kinematic constraints without doing work on admissible virtual velocities. Indeed, for any $v\in V$,
\[
\int_\Omega v\cdot \nabla P_R\,dV
= \int_{\partial\Omega} P_R\,v\cdot n\,dS - \int_\Omega P_R\,\nabla\!\cdot v\,dV = 0.
\]
Thus the reaction pressure is $L^2$-orthogonal to the space of admissible velocities, exactly as in d'Alembert's principle, and the ``multiplier'' is most naturally interpreted as the gradient component required to project $\rho(u_t+C)$, rather than as a finite-dimensional rank count.

\rev{One additional point is that, unlike in finite-dimensional vector spaces, norms are not equivalent in the function-space setting. Physically, the advective term  $(u \cdot \nabla)u$ is  unlikely to remain an $L^2$ function in most Euler solutions. Thus, other weaker norms such as the $H^1$-norm are more plausible and have been considered in related works such as ~\cite{hoffman2010resolution} and ~\cite{hoffman2016new} and offer room for additional explorations.}

\end{document}